\documentclass[twocolumn,
reprint,
aip,
pop,
superscriptaddress,
floatfix]{revtex4-1}
\usepackage[utf8]{inputenc}
\usepackage[T1]{fontenc}
\usepackage[english]{babel}

\usepackage{amsmath}  
\usepackage{amsfonts}
\usepackage{amssymb}
\usepackage{amsthm}

\usepackage{textcomp}

\usepackage{graphicx}

\usepackage{color}

 \usepackage{hyperref}
\hypersetup{
    colorlinks=true,
    citecolor=blue,
    linkcolor=blue,
    filecolor=blue,      
    urlcolor=blue,
}

\usepackage{fancyhdr}
\begin{document}

\title{Electromagnetic pulse emission from target holders during short-pulse laser interactions}

\author{Damien~F.~G.~Minenna}
\altaffiliation{Present address: CEA, DAM, DIF, F-91297 Arpajon
Cedex, France.}
    \affiliation{%
    Centre National d'{\'E}tudes Spatiales, 31401 Toulouse cedex 9, France
    }%
    \affiliation{%
    Aix-Marseille Universit{\'e}, CNRS, PIIM, UMR 7345, 13397 Marseille, France
    }%
    \affiliation{%
    Thales AVS, 78140 V{\'e}lizy, France
    }%
\author{Alexandre~Poy{\'e}}%
    \email[Corresponding author:\\]{alexandre.poye@univ-amu.fr}
    \affiliation{%
    Aix-Marseille Universit{\'e}, CNRS, PIIM, UMR 7345, 13397 Marseille, France
    }%
\author{Philip Bradford}
    \affiliation{Department of Physics, York Plasma Institute, University of York, Heslington, United Kingdom}    
\author{Nigel Woolsey}
    \affiliation{Department of Physics, York Plasma Institute, University of York, Heslington, United Kingdom}   
\author{Vladimir~T.~Tikhonchuk}
    \affiliation{Centre Lasers Intenses et Applications, University of Bordeaux-CNRS-CEA, 33405 Talence, France}
    \affiliation{ELI-Beamlines, Institute of Physics, Czech Academy of Sciences, Za Radnicic 835, 25241 Dolní Břežany, Czech Republic}


\begin{abstract}
For the first time, a global model of electromagnetic pulse (EMP) emission connects charge separation in the laser target to quantitative measurements of the electromagnetic field. We present a frequency-domain dipole antenna model which predicts the quantity of charge accumulated in a laser target as well as the EMP amplitude and frequency. The model is validated against measurements from several high-intensity laser facilities, providing insight into target physics and informing the design of next-generation ultra-intense laser facilities. EMP amplitude is proportional to the total charge accumulated on the target, but we demonstrate that it is not directly affected by target charging time (and therefore the laser pulse duration) provided the charging time is shorter than the antenna characteristic time. We propose two independent methods for estimating the charging time based on the laser pulse duration. We also investigate the impact of target holder geometry on EMPs using cylindrical, conical and helical holders. 
\end{abstract}

\maketitle
\thispagestyle{fancy}

\noindent Published as: Minenna \textit{et al.}, Phys.\ Plasmas, \textbf{27}, 063102 (2020),  \href{https://doi.org/10.1063/5.0006666}{https://doi.org/10.1063/5.0006666}.

\section{Introduction}

Addressing the problem of electromagnetic pulse (EMP) emission is key to taking advantage of recent
developments in high-repetition, high-intensity laser facilities.\cite{dan19} During and after laser shots on solid targets, intense EMPs are produced inside the laser area. 
This phenomenon has been known since the seventies and has been investigated at numerous facilities.\cite{pea78,mea04,rai04,sto06,rem07,bro08,ede09,bat12,bro13}
Laser-driven EMP emission, spanning a broad frequency range from MHz to THz, can seriously disrupt electronic equipment used for facility operation or scientific measurement. On the other hand, if properly controlled, these emissions may lead to new experimental applications.\cite{Kar16,wang18} Investigating sources of EMP is a challenge because the phenomenon is not limited to a single physical process.\cite{conPreprint} EMP in the THz domain is generated from electron oscillations in the target and is characterized by the duration of electron ejection.\cite{lia19,liu18,liu19}
Simultaneously, electromagnetic fields propagating through an interaction chamber will activate all the metallic parts, emitting EMP at lower frequencies ($\sim{}$100~MHz) as chamber proper modes. 

In this paper, we focus on EMPs in the GHz domain. During and after a laser shot, hot electrons are
ejected from the target, leaving behind a net positive charge. The total accumulated charge produces
a return current (i.e.\ neutralization current) that propagates along the target holder to the ground and
induces a GHz EMP.\cite{conPreprint,qui09,dub14,cik14,poy15,san15,con16,rac17,rob17,tik17,kra17,bra18}
The combined target-target holder system acts as an antenna
composed of a capacitance (target) and an inductance (holder). EMP emission from this system is governed by two characteristic times: the charging time of the target (electron ejection period) and the antenna time (oscillation period of the return current). 

\begin{figure}
    \centering
    \includegraphics[width=\columnwidth]{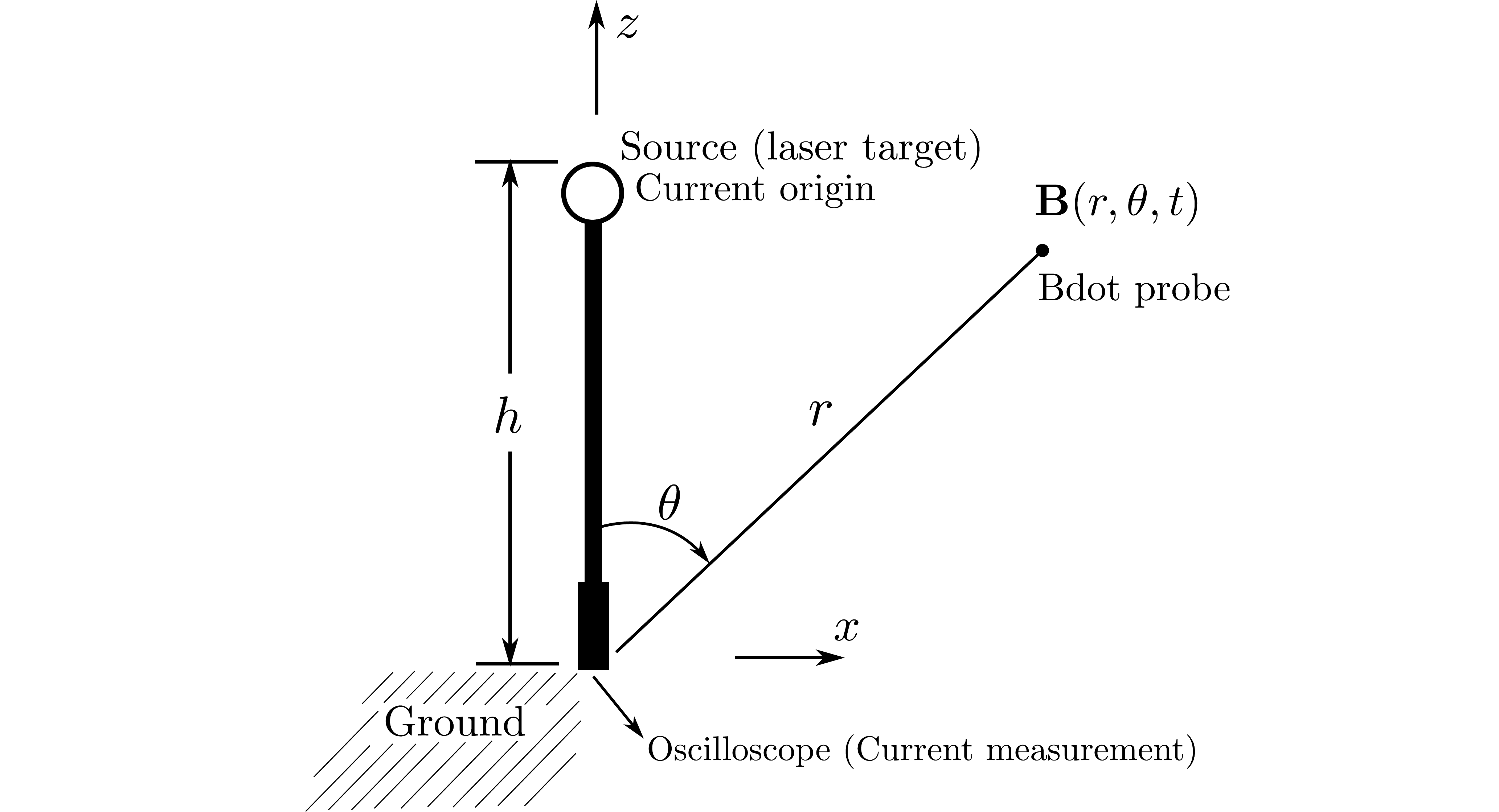}
    \caption{\label{fig:scheme} Schematic of the experimental set-up. The bold line represents the target holder (see Fig.~\ref{fig:holder} for the various holder designs used). During a laser shot, charge accumulation in the target generates a return current that propagates down the target holder. The magnetic field is measured by a B-dot probe and the neutralization current is recorded using an oscilloscope connected by a coaxial cable to the base of the holder.} 
\end{figure} 

This paper builds on the theoretical model described in Ref.~\onlinecite{poy15} and the associated erratum to describe the
magnetic field. We test the model against a wider parameter
space, including a much expanded laser energy
range and different target holder geometries. These developments illustrate how our frequency-domain dipole
antenna model gives a good qualitative description of
EMP emission, providing order-of-magnitude estimates
of magnetic flux, or, alternatively, an estimate of the positive
charge accumulated on a target based on magnetic
field measurements.

This model is compared to experimental data from the ECLIPSE and Vulcan facilities. We show that
our model is accurate provided the charging time is shorter than the antenna time. Then we compute
the target charging time using two different methods and compare them numerically. 

Our paper is is organized as follows: Sec.~\ref{sec:expSetup} contains the experimental set-up and methodology, then
spectral analysis of the neutralization current and EMP from different target holders is presented in
Sec.~\ref{sec:measurements}. In Sec.~\ref{sec:frequency}, we present the antenna model and test it against measurements. Finally, in Sec.~\ref{sec:charging}, we compute the target charging time. The main results are summarized in Sec.~\ref{sec:conclu}.

\section{Experimental set-up}
\label{sec:expSetup}

\begin{figure}
    \centering
    \includegraphics[width=\columnwidth]{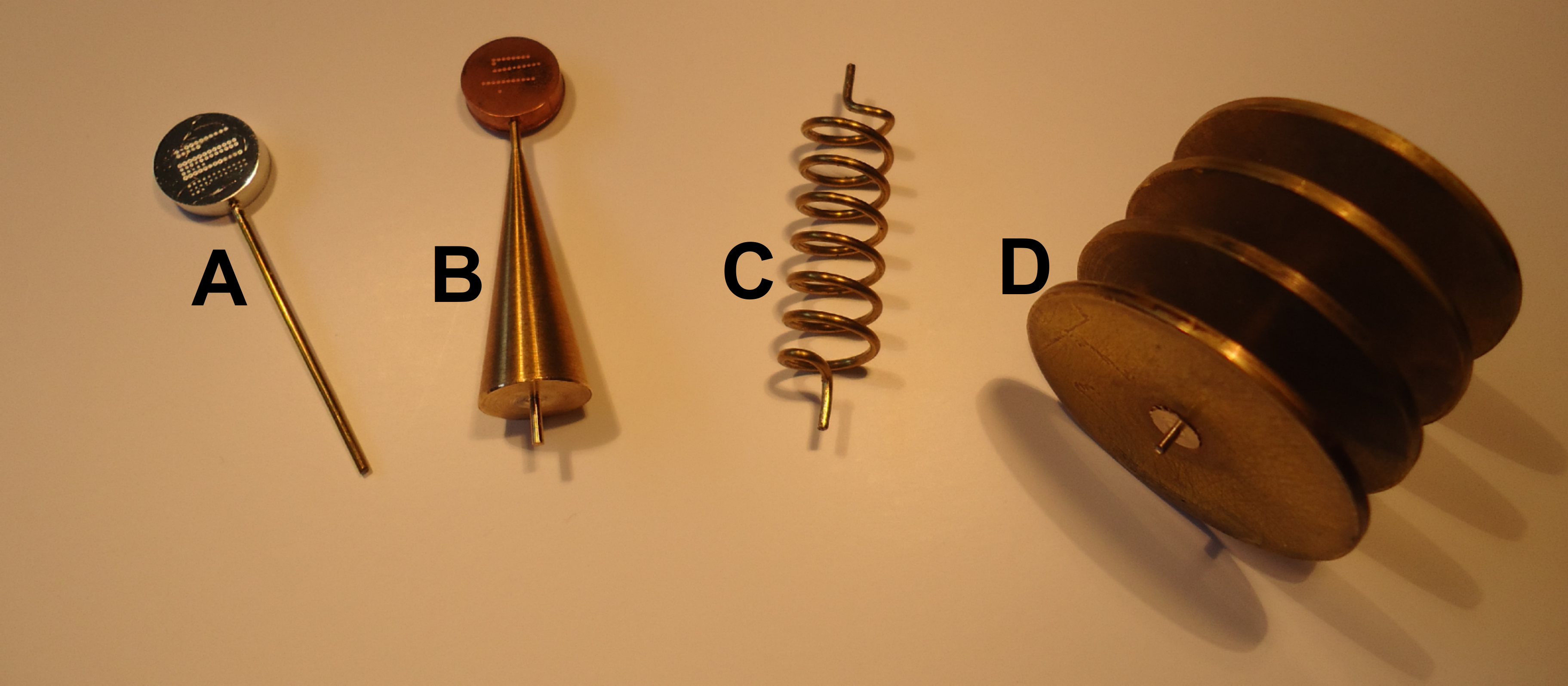}
    \caption{\label{fig:holder} 
    (Color online) Photo of different target holder designs, labelled A, B, C and D. The laser target (not displayed on C and D) is a metallic disk 3~mm thick, $d=1$~cm in diameter, made from aluminium, copper or tantalum. Holder A is a 1~mm-diameter conducting wire with a length (distance between the target and the ground plate) of $l=4.5$~cm. For a detailed description of the experiment, see Refs \onlinecite{dub14,poy15}.} 
\end{figure} 

Our experiment was conducted at the ECLIPSE laser facility. The set-up is fully described in Fig.~\ref{fig:scheme} and Refs \onlinecite{dub14,poy15,rac17}. A Ti:Sapphire laser with wavelength 807~nm and maximum intensity $10^{18}$~W/cm$^2$ was used to drive EMP emission from a metallic target. Laser energy $E_{\rm las}$ was varied from 10 to 100~mJ and focused to a radius of $r_{\rm las} = 6~\mu$m.
The intensity contrast is around $10^{-7}$, which corresponds to a laser absorption efficiency of 40~\%.
The laser pulse duration $t_{\rm las}$ ranged from 0.03 to 10~ps. Simultaneous measurements were made of the return current and magnetic field using respectively a 2~m-long coaxial cable connected to the base of the target holder and a B-dot probe. 
Both the neutralization current and the magnetic field were recorded during and after the laser shot.
Current measurements were taken using a 6~GHz bandwidth oscilloscope with an attenuation of 60~dB at the scope output. Effects of the cable and attenuators were removed from the data during post-processing.
Integration of the current profile provides the experimental total target charge.
The position of the B-dot probe for all measurements was $r=240$~mm and $\theta=77$\textdegree.
Fourier analysis of the EMP signal was integrated over the whole signal duration (10002 points spaced by 50 ps). We used the window proposed by Blackman-Harris over 1000 points, with a hop width (space between each short-term FT) of 30 points. The oscilloscope is limited to 7~GHz.
A 3D simulation of EMP emission inside the ECLIPSE chamber is available as a video in the supplementary material of Ref.~\onlinecite{poy15}.

The bold line in Fig.~\ref{fig:scheme} represents the target holder. Figure~\ref{fig:holder} details the various target holders used in our experiment. Each holder was fixed to a connector in the middle of a large ($20\times20$~cm$^2$) metallic ground plane. The size of this ground plane was chosen deliberately so that it would behave as an electromagnetic mirror and the target holder would emit as a dipole antenna (see below).
Holder A is a straight wire, 1~mm in diameter, with a length $l=4.5$~cm. Holder C is a helix with $n=7$ periods, a pitch $p_{\rm h} = 0.57$~cm and a diameter $d_{\rm h} =1$~cm.
The targets are metallic disks, $d=1$~cm in diameter and 3~mm thick, made from aluminium (Al), copper (Cu) or tantalum (Ta).
Since the targets themselves contribute towards the antenna emission, we approximate the length of holder A to $h \approx l + \pi d /2$ in our models.\cite{poy15} Note that this experimental arrangement is similar to Ref.~\onlinecite{bra18}, where different holder designs were also investigated.

\begin{figure}
    \centering
    \includegraphics[width=\columnwidth]{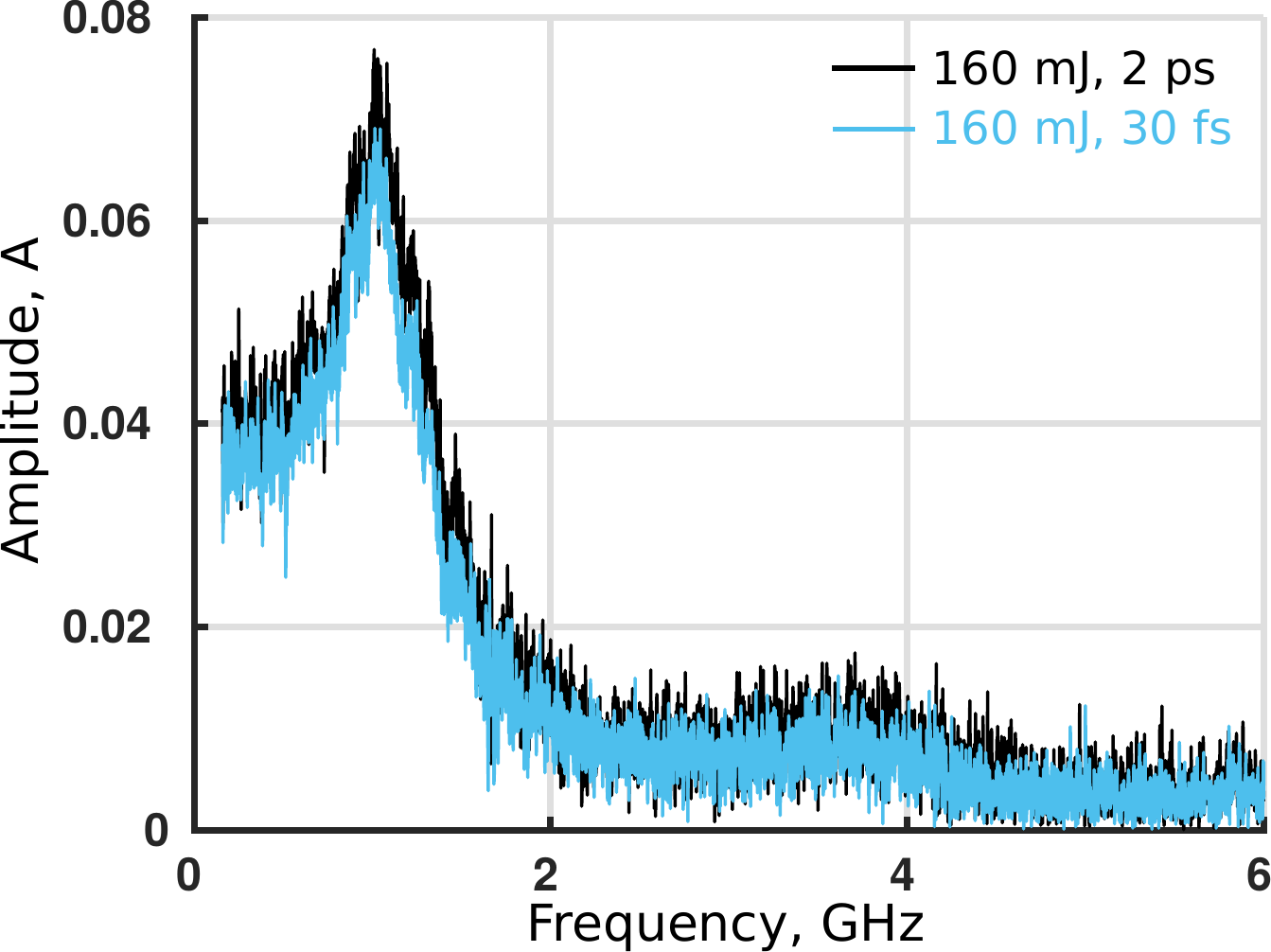}
    \caption{\label{fig:Ispectr}(Color online) Spectrum of the neutralization current for two laser shots using holder A. 
    Light blue curve: $E_{\rm las}=160$~mJ, $t_{\rm las}=30$~fs.
    Black curve: $E_{\rm las}=160$~mJ, $t_{\rm las}=2$~ps.
    For both plots, the resonant frequency is 1~GHz with a bandwidth of $\pm 0.3$~GHz.} 
\end{figure}

\begin{figure*}
\centering
\includegraphics[width=\textwidth]{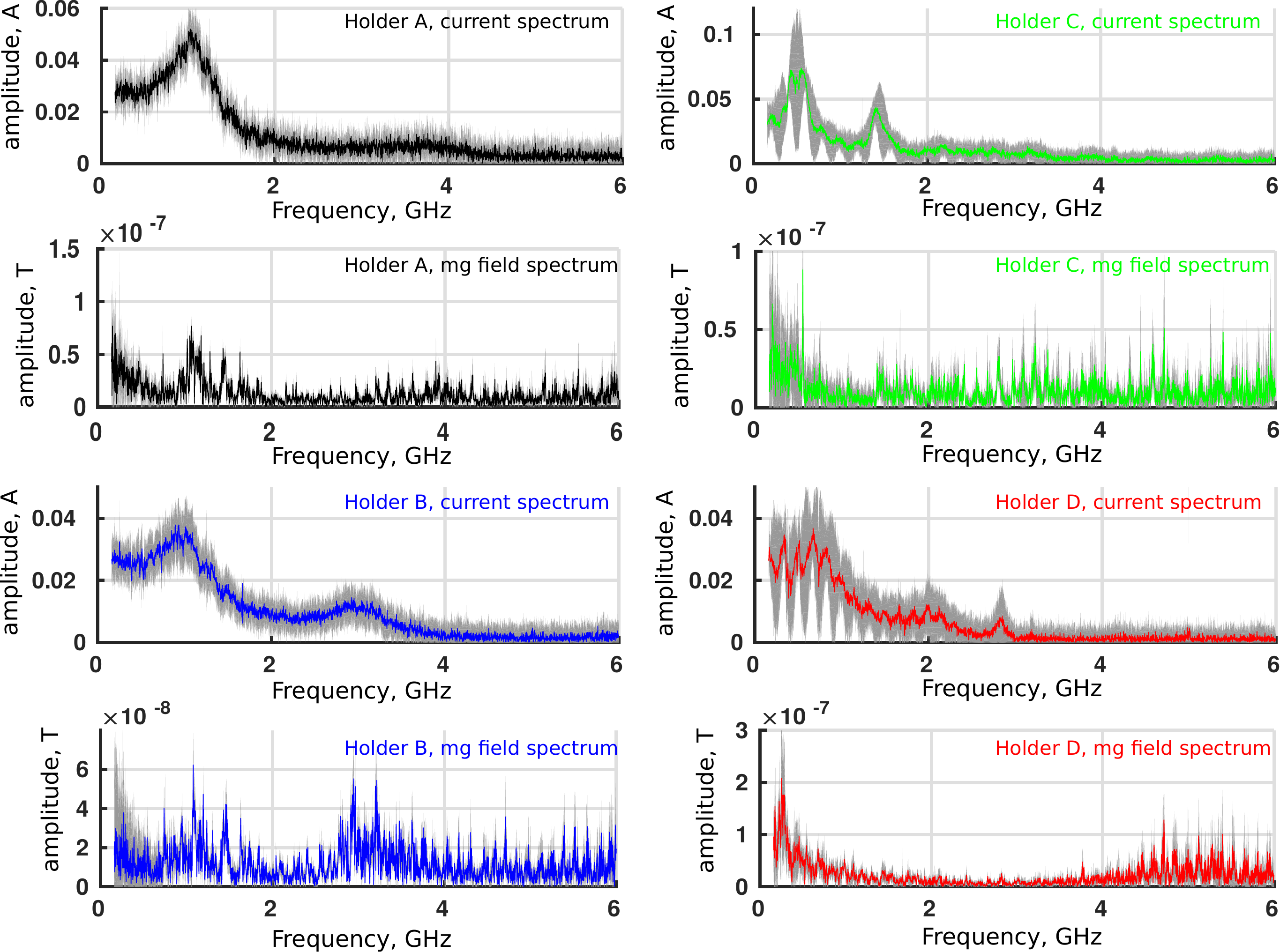}
 \caption{(Color online) Measured spectra of the return current (top panel) and emitted magnetic field (bottom panel) for different target holders (see Fig.~\ref{fig:holder}). 
 Colored curves are averaged over 10 shots.
 The grey shadows are the standard deviation around the measurement.
 There is a correlation between the resonant frequencies in the current and in the magnetic field. Laser parameters were identical on all shots with $E_{\rm las}=120$~mJ and $t_{\rm las}=30$~fs. The targets were identical copper disks.} \label{fig:spectrum_holder}
\end{figure*} 

According to Ref.~\onlinecite{pea77}, the return current forms a transient pulse because the antenna is not fed continuously as it radiates. First, charge is collected on the target during the hot electron ejection process.\cite{dub14,poy15}
The ChoCoLaTII.f90 model\cite{poy15b,poy18} predicts that, at the level of the target ($z=h$), the current has a Gaussian distribution with a standard deviation related to the charging time, $t_{\rm e}$.
As it propagates down the target holder, this current pulse is stretched because the holder behaves as a distributed inductance.\cite{ahm17,akt19} The antenna functions as a low-pass frequency filter, removing wavelengths larger than the dipole wavelength, $\lambda_\tau$. The shorter the neutralisation pulse, the more it will be stretched out over the antenna surface. Long ($\sim{}$ns) neutralisation pulses can be considered as a constant current supply for cm-scale antennas, producing weak EMPs in the GHz domain. This defines a condition for efficient EMP generation: $t_{\rm e} \ll \tau$, where $\tau = 1 / f_\tau$ is the time period of the emission. For holder A, we have $\tau = 4 h /c$.
Further on, we consider both times: the antenna time $\tau$ is used in Sec.~\ref{sec:frequency} to estimate the maximum EMP magnetic field and the charging time $t_{\rm e}$ is estimated in Sec.~\ref{sec:charging}.

\section{Impact of target holder on EMP and current spectrum}
\label{sec:measurements}

\subsection{Return current spectrum}

\begin{figure*}
    \centering
    \includegraphics[width=\textwidth]{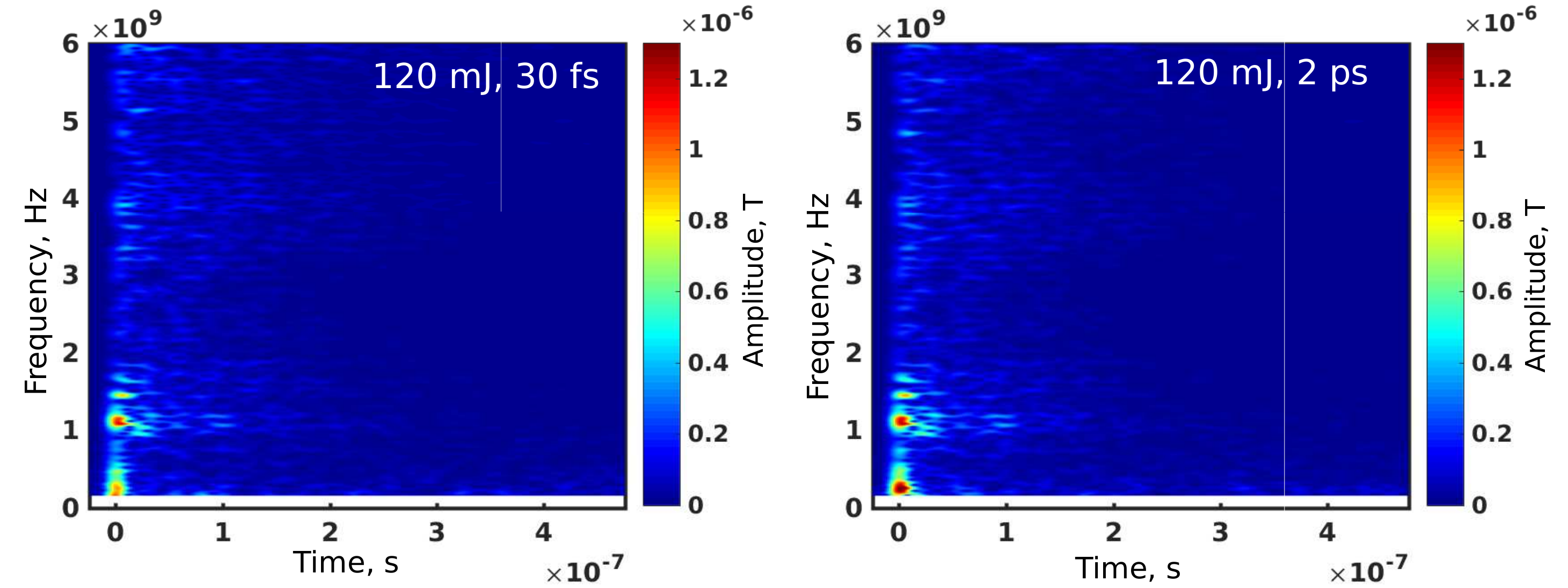}
    \caption{\label{fig:spectrogram}
    (Color online) Time-dependent spectrograms of the emitted magnetic field for target holder A. The EMP was measured using B-dot probe positioned at $r=240$~mm and $\theta=77$\textdegree. For both plots, a laser energy is $E_{\rm las} =120$~mJ and a focal radius is $r_{\rm las}= 6$~$\mu$m.
    Left panel: Pulse duration of $t_{\rm las} =50$~fs.
    Right panel: Pulse duration of $t_{\rm las} =2$~ps.}
\end{figure*}

Target holders shaped like straight wires (see holder A) will radiate like a ground plane (dipole) antenna.\cite{hea95,jac99}
If the metallic ground is large enough, it acts as a mirror and the length of the target holder is equal to the quarter-wavelength of emission $h = \lambda_\tau / 4$.
This was demonstrated in Ref.\ \onlinecite{poy15}, when spectral analysis of the neutralization current and magnetic field revealed a strong resonance corresponding to the the antenna frequency $f_\tau = 1 / \tau = c / (4 h)$ or to the wave-number $k_\tau = \pi / (2 h)$.
Figure~\ref{fig:Ispectr} shows the spectrum of the neutralization current for two laser shots on target holder A with different laser energies and pulse durations. The position of the spectral peak coincides with the antenna frequency $f_\tau$ for both shots, even though the laser pulse duration differs by a factor of $\sim$70.
Without a ground, the half-wavelength of emission should be $\lambda_\tau = 2 h$.

The current spectrum also depends on the shape of the target holder.
The top plots in Fig.~\ref{fig:spectrum_holder} show how the spectrum is modified for different target holders, measured from the base of the holder with identical laser parameters. 
The holder A spectrum consists of a single spike around $f_\tau = 1.2$~GHz.
For holder~B, we observe the addition of a third harmonic at $3$~GHz compared to holder~A. Holders~C and D both have a lower resonant frequency.

\subsection{Magnetic field spectrum}

Figure~\ref{fig:spectrum_holder} also shows spectra of the magnetic field measured by the B-dot probe (bottom panels). Although the spectra are more noisy, we recover the same resonant frequencies for holders A and B. 
However, we note that the third harmonic in holder B has a stronger amplitude in the magnetic field than in the current. 
Holder C has a common peak in the current and the magnetic field at 0.5~GHz, while holder D has a higher amplitude at its main frequency ($f_\tau = 0.3$~GHz) than the other holder designs.
For each holder, there are artefacts associated with their particular design.
Within the dipole model (described below), the antenna is supposed to be a thin straight wire with a perfect metallic ground. 
But our holders are composed of three different parts (the ground, the holder and the target) that alter the model prediction.
For instance, the target itself contributes towards the emitted spectrum, with resonances between 3 and 5~GHz according to its size. We have no information about the directivity of the EMP emission because measurements were taken with a single B-dot probe, although other publications confirm the dipolar emission pattern.\cite{xia20}

Figure~\ref{fig:spectrogram} shows spectrograms for two EMPs emitted from targets suspended on A-type holders, with $t_{\rm las} = 50$~fs (left plot) and $t_{\rm las} = 2$~ps (right plot). All other laser parameters are identical. 
We observe that both plots are qualitatively the same, despite the laser intensity downgrading by two magnitude orders. 
This is consistent with Ref.~\onlinecite{poy15}, which shows that the total accumulated charge, $Q$, is independent on the laser pulse duration for $t_{\rm las} = 30$~fs to 5~ps, based on data from the same experiment with energies around 100~mJ. If $Q$ is the same for both shots, this would explain the similar current spectra in Fig.~\ref{fig:Ispectr}.
In Sec~\ref{sec:frequency}, we argue that the EMP amplitude is proportional to $Q$. Thus, at low laser energy and short pulse duration, EMP spectra are unaffected by the laser pulse duration or the charging time (since the target charging time is related to the laser pulse duration - see Sec.~\ref{sec:charging}). At higher laser energies, $Q$ can become dependent\cite{akt19} on $t_{\rm las}$, so the EMP amplitude might therefore be affected (see Fig.~\ref{fig:Brad_Elas} below).

If the reader refers to Fig.~\ref{fig:spectrogram}, it is possible to identify the EMP resonant frequencies:
Signal around 1~GHz is related to the antenna (holder A). Frequency content below 0.2~GHz is background signal from the experimental chamber that persists long after the laser pulse has ended. The remainder of the signal is in the 3 to 6~GHz range and isn't present in the return current (see Fig.~\ref{fig:Ispectr}). This implies emission from millimeter-sized objects: probably the target itself.
The target dimensions (1~cm diameter and 3~mm thickness) are in a good range to produce the EMP signal from 3 to 6~GHz. The charging time, $t_{\rm e}$, is still short enough to satisfy the condition of efficient EMP generation for our values of the laser energy and pulse duration.

\begin{figure}
    \centering
    \includegraphics[width=\columnwidth]{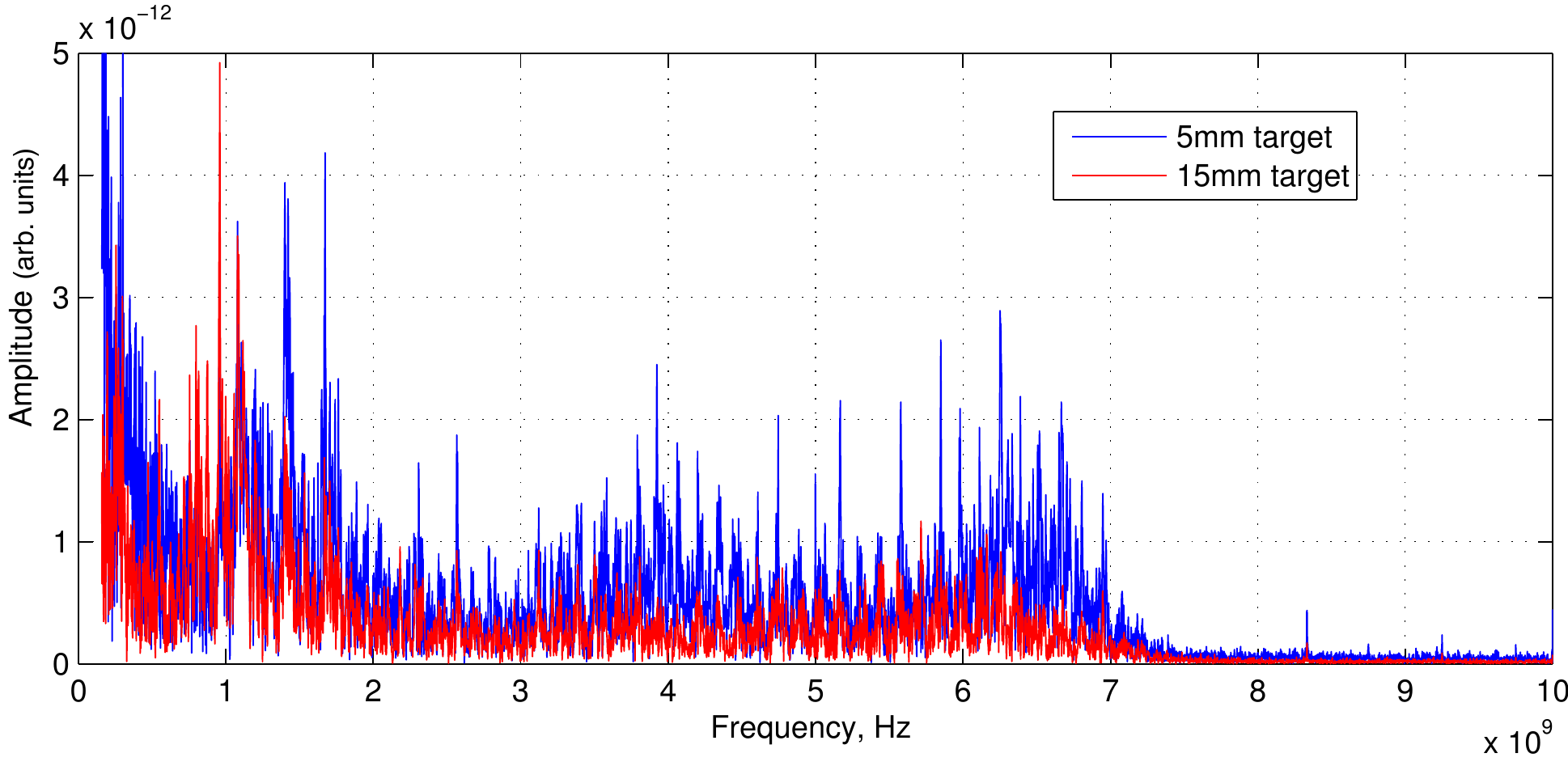} 
    \caption{(Color online) Comparison of magnetic field spectra for two target diameters $d=5$~mm and $d=15$~mm. Both targets were mounted on A-type holders and the laser parameters are identical.} \label{fig:spectrogram_2sizes}
\end{figure}

To illustrate the impact of target size on the EMP spectrum, we compare two targets with diameters $d= 5$ and $d=15$ mm (data is taken from the campaign reported in Ref.~\onlinecite{dub14}).
Figure~\ref{fig:spectrogram_2sizes} shows differences in the magnetic field spectrum. 
The larger target, reciprocally the smaller target, shifts the spectrum to lower frequencies, reciprocally to higher frequencies. The shift of the 1~GHz peak is related to the holder size, including the target diameter. At the same time, increasing the target size reduces the signal in the 3 to 6~GHz range. This damping might be related to the coupling between the global antenna of height \textit{h} and oscillations in the target itself. However, determination of the target eigenfrequency is not investigated here.

\section{Estimating the magnetic field amplitude of EMP}
\label{sec:frequency}

\subsection{Frequency-domain model}

Estimating the maximum amplitude of the magnetic field prior to a laser shot is crucial for the protection of electronic equipment. We propose a method to compute it as a function of the total charge ejected by the laser. This method is based on a classical dipole antenna model in the frequency-domain.\cite{hea95,jac99}


The target holder is assumed to be a thin metallic stalk of height $h$, similar to holder A. We also assume the charging time is much less than the antenna time ($t_{\rm e}  \ll \tau = 4h /c$), so the target-holder system is qualitatively equivalent to a straight wire of length $2h$, with point charges $+Q$ and $-Q$ attached to each end. These point charges will oscillate along the virtual wire and emit radiation at the antenna frequency, $f_\tau = c/(4h)$. Under these conditions, a dipole approximation is valid
and EMP emission will be maximal.

At the antenna wave-number $k_\tau$ (equivalently, the antenna frequency), the magnitude of the magnetic field for a dipole antenna in the far-field region is given by
\begin{equation}
    | {\bf B}_{k_\tau} (r,\theta) | = \frac{\mu_0 \left|\tilde I_{k_\tau} \right|}{2\pi r} \left| \frac{\cos\left( (\pi/2) \cos\left(\theta\right)\right) }{\sin\theta} \right| \, , \label{eq:Bquaterantenna}
\end{equation}
 where $\tilde I_{k_\tau}$ is the antenna current at the antenna characteristic frequency, $r$ is the radial distance measured from the base of the antenna, $\theta$ is the angle with respect to the antenna axis and the oscillation wavelength satisfies $h \ll \lambda \ll r$. To use Eq.~\eqref{eq:Bquaterantenna} in our model, we must take the neutralization current to be a sinusoidal function of time. This approximation is crude but sufficient for an estimation of the maximum magnetic field. The maximum return current at the antenna frequency can be expressed as
\begin{equation}
    \tilde I_{k_\tau} = Q f_\tau \, , \label{eq:currentItauA}
\end{equation}
with $Q$ the total charge and $f_\tau$ the antenna frequency. The value of this frequency can be estimated from the length of the target holder. For example, \ $f_\tau = c / (4 h)$ for holder A, or, if the shape of the antenna is more complicated, it can be directly measured in a preliminary experiment at low energy.

\begin{figure}
\centering
\includegraphics[width=\columnwidth]{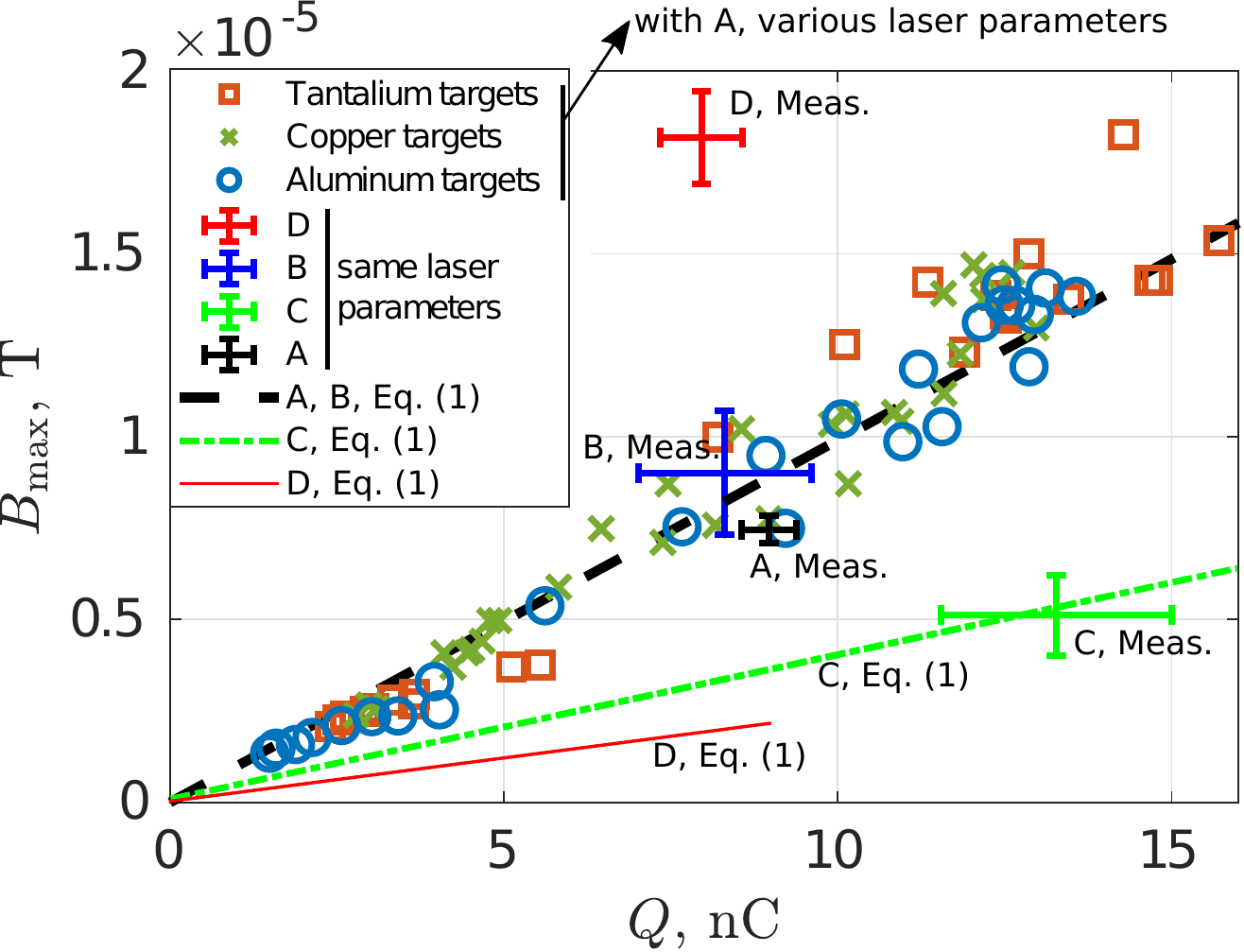}
\caption{\label{fig:Q_B} (Color online)
Peak magnetic field, $B_{\rm max}$, plotted as a function of the target charge for different laser parameters, target holders and targets. All individual data points represent the average of 5 experimental shots on the ECLIPSE laser facility.\cite{dub14}
Straight lines are calculated using Eq.~\eqref{eq:Bquaterantenna}, with $f_\tau$ estimated from the holder dimensions. Good agreement is seen between experimental measurements and the frequency-domain model for holders A, B and C.
The small crosses, circles and squares are taken from Ref.~\onlinecite{poy15}. These targets were mounted on target holder A for laser parameters $E_{\rm las} =10$--$80$~mJ,  $t_{\rm las} = 0.03$--$10$~ps and $r_{\rm las}= 6$~$\mu$m, with 40\% estimated laser absorption. The targets are 3~mm-thick disks, $1$~cm in diameter, made from aluminium (circles), copper (crosses) or tantalum (squares). The points with error bars are identical laser shots using four different holder designs (A, B, C and D). The black dashed line is Eq.~\eqref{eq:Bquaterantenna} with $f_\tau = 1.23$ GHz, valid for holders A and B.
The green dot-dashed line assumes $f_\tau = 0.5$ GHz, valid for holder C, while the red line (plotted only between $Q = 0$ and $9$~nC) takes $f_\tau = 0.3$ GHz, valid for holder D.} 
\end{figure}

Equation~\eqref{eq:Bquaterantenna} has previously been used to determine the magnetic field produced by a laser-target holder (see Ref.\ \onlinecite{poy15}). However the current was incorrectly assumed to be $\tilde I_{k_\tau} = 1.6 \, Q c / h$ instead of Eq.~\eqref{eq:currentItauA}, thus it was 6.4 times higher. This error went undetected because there was a missing $1/(2 \pi)$ factor in the measurement data. This $1/(2 \pi)$ factor has since been corrected in the associated erratum.\cite{poy15}
In summary, we propose Eq.~\eqref{eq:currentItauA} as a more accurate description of the maximum current flowing in the target.

Combining equations~\eqref{eq:Bquaterantenna} and~\eqref{eq:currentItauA}, one can easily evaluate the order of magnitude of EMPs generated by a neutralization current, particularly if the holder is a thin metallic stalk (e.g.\ holder A) with a large metallic ground. 
The total charge $Q$ can be quickly estimated, for example, with the ChoCoLaTII.f90 code.\cite{poy15b,poy18}
In addition, this expression provides us with a practical method to control the EMP amplitude: For target holders shaped like a straight wire, the magnetic field is directly proportional to the antenna frequency and therefore inversely proportional to $h$. To decrease the EMP amplitude, one can simply increase the effective stalk length.\cite{bra18}

\begin{figure*}
\centering
\includegraphics[width=\columnwidth]{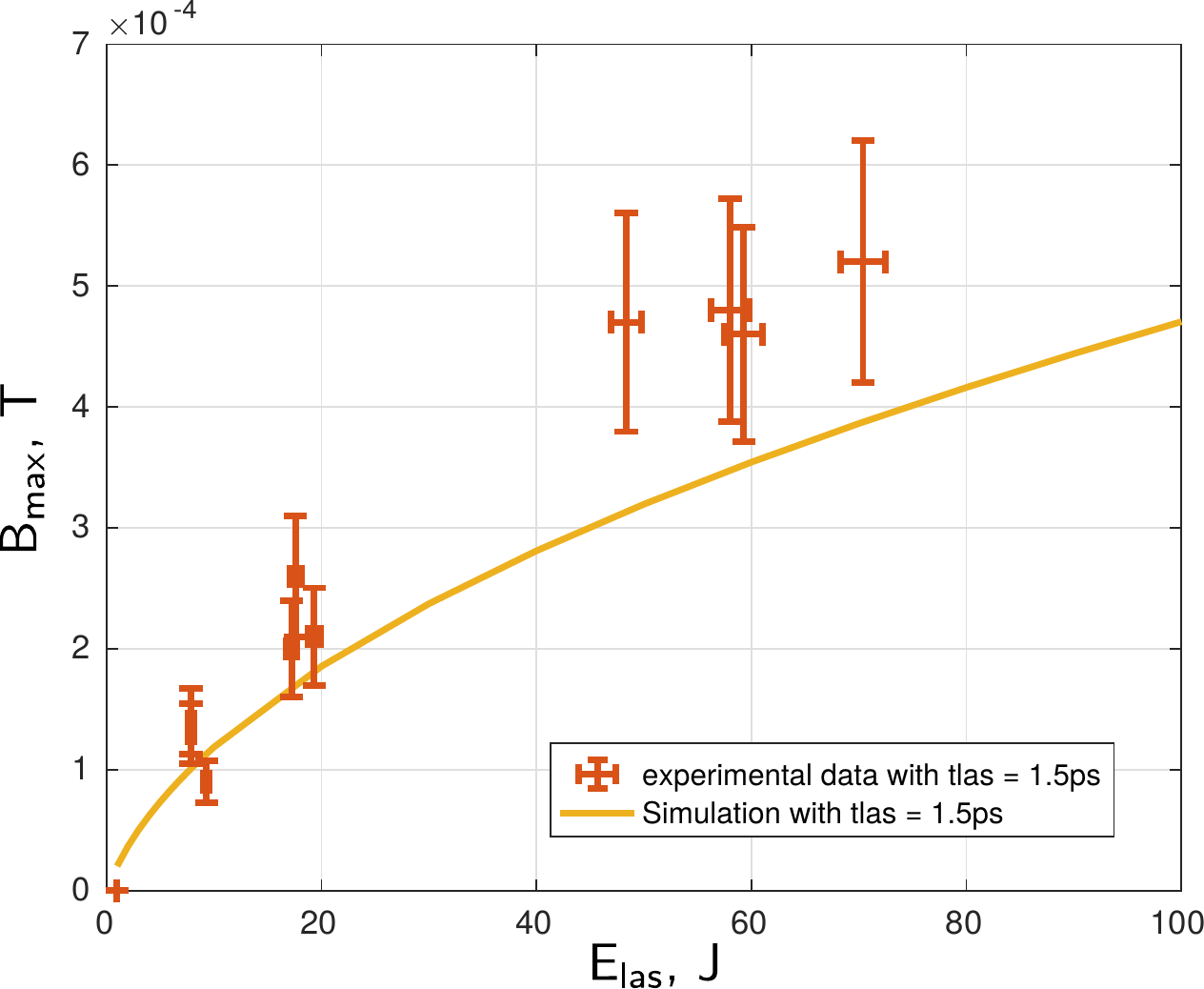} \includegraphics[width=\columnwidth]{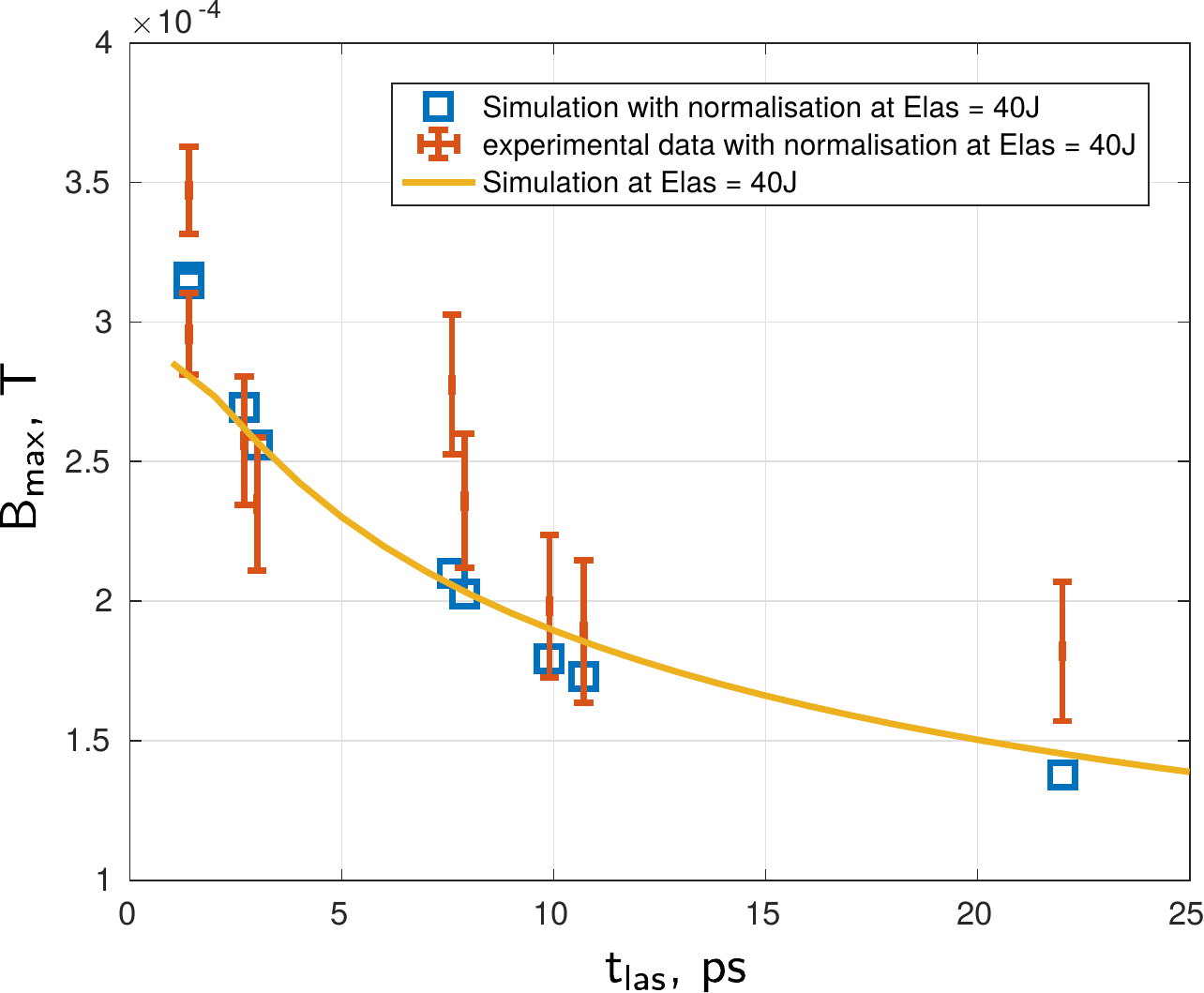}
\caption{\label{fig:Brad_Elas} (Color online)  Maximum magnetic field plotted as a function of the laser energy (left plot) and the laser pulse duration (right plot). The points with error bars represent experimental data from the Vulcan laser facility.\cite{edw98} Yellow lines and blue squares are from our model Eq.~\eqref{eq:Bquaterantenna} at $r = 1.5$~m and $\theta=90$\textdegree{}. $Q$ is evaluated from the model ChoCoLaTII.f90.} 
\end{figure*}

\subsection{Comparison of model with ECLIPSE data}

To assess Eq.~\eqref{eq:Bquaterantenna}, we performed several experimental measurements as described in Sec.~\ref{sec:measurements} using the ECLIPSE laser facility.\cite{dub14}
Figure~\ref{fig:Q_B} displays the maximum magnetic field recorded by the B-dot probe as a function of the accumulated charge. The small crosses, circles and squares represent data taken from Ref.~\onlinecite{poy15} (see the associated erratum). Laser shots were performed for various laser energies and laser duration times using targets fashioned from aluminium, copper or tantalum and mounted on target holder A. For these shots, the maximum magnetic field is influenced only by the accumulated target charge. Predictions from the frequency-domain model (Eq.~\eqref{eq:Bquaterantenna}) are plotted as straight lines. We see excellent agreement between the frequency-domain model and measurements for holder A because it is close to an ideal dipole antenna. The model should be applied with care to holders of other shapes, however.
In Fig.~\ref{fig:Q_B}, we have also plotted the maximum magnetic field recorded for four different holder designs (A, B, C and D) under the same experimental conditions. Our model, at $f_\tau = 1.23$ GHz,  predicts a near-identical peak magnetic field for holders A and B because their physical size and Fourier spectrum are similar. 
Discrepancies between the model at $f_\tau = 1.23$ GHz and experiment are stronger for holders C and D.
According to Fig.~\ref{fig:spectrum_holder}, the resonant frequency of holder C is $f_\tau = 0.5$ GHz, which corresponds well to the length of its wire, $l = n (\pi^2 d_{\rm h}^2 + p_{\rm h}^2)^{1/2}$.
Similarly, based on the measured spectra, the resonant frequency of holder D is $f_\tau = 0.3$ GHz.
Choosing the appropriate frequency in Eq.~\eqref{eq:Bquaterantenna}, we see that our model works well for holder C but not for holder D. The implication is that holder D cannot be modeled as a quarter wavelength antenna with a characteristic length, $h$.

The experimental data suggests that, for the same laser parameters and targets, the EMP amplitude is two times larger with holder D compared with holder A and B, while it is more than two times lower with holder C. Thus a helical stalk can be used to significantly reduce the EMP field amplitude.\cite{conPreprint,bra18}

\subsection{Comparison of model with Vulcan data}

We provide a further test of our model using data from Ref.~\onlinecite{bra18}. The experiment was performed on the Vulcan laser facility,\cite{edw98} with a maximum intensity of $2 \times 10^{19}$~W/cm$^2$. The laser energy was varied from $E_{\rm las} = 10$ to 70~J and the laser pulse duration from $t_{\rm las} = 1$ to 22~ps. The laser intensity contrast\cite{Mus10} was about $10^{-8}$. Magnetic field measurements were taken at $r = 1.5$~m and $\theta=90$\textdegree{}. The antenna frequency is $f_\tau = 2.9$~GHz based on the height of the target holder. The total charge on the target is evaluated using ChoCoLaTII.f90 according to the experimental laser parameters. The electron ejection time, $t_{\rm e}$, is much smaller than the antenna characteristic time, so we can apply Eq.~\eqref{eq:Bquaterantenna} to get the maximum magnetic field. The maximum magnetic field (both measured and simulated) is plotted in Fig.~\ref{fig:Brad_Elas} as a function of the laser energy $E_{\rm las}$ and separately as a function of the laser pulse duration, $t_{\rm las}$. Since the laser energy was not constant during the pulse duration scan, the Vulcan data has been normalized to a reference energy of 40~J using a linear fit to the energy scan data (red crosses in the left plot of Fig~\ref{fig:Brad_Elas}). The blue squares represent simulations run with the experimental values of the laser energy that have then been normalised to 40~J using the linear fit. To confirm the validity of this normalization process, we also plot a simulated pulse duration scan at fixed energy $E_{\rm las} = 40$~J (yellow curve in right plot). There is a reasonable agreement between the measurements (red points with error bars) and our simulations (yellow lines and blue squares).


\subsection{Estimating the target charge from magnetic field measurements}

The target charge, $Q$, is ideally calculated from experimental measurements of the return current. However, our frequency-domain model can also be used to estimate $Q$ from measurements of the magnetic field. Consider an experiment at the LULI2000 laser facility\cite{luli} (see Ref.\ \onlinecite{conPreprint}). The experiment was performed with $E_{\rm las} = 80$~J, $t_{\rm las} = 1.3$~ps and $\sim{}10~\mu$m FWHM, for an overall intensity $10^{19}$~W/cm$^2$. The intensity contrast was about $10^{-6}$ on this facility.\cite{Puy19} A B-dot probe was positioned at $r = 54$~cm and $\theta=90$\textdegree{}. The probe measured a peak magnetic field $| B_{\rm max} | = 1.5 \times 10^{-4}$~T and resonant frequency $f_\tau = $ 1~GHz. This resonant frequency corresponds to $c/(2 h)$ - the factor 2 instead of 4 because there is no ground (mirror) in the experiment. Integration of the current measurements gives the total charge of approximately $Q =  270$~nC. According to Eqs.~\eqref{eq:Bquaterantenna} and \eqref{eq:currentItauA}, the charge should be $Q =  407$~nC - just 1.5 times larger than in the experiment. 

\begin{figure*}
\centering
\includegraphics[width=\columnwidth]{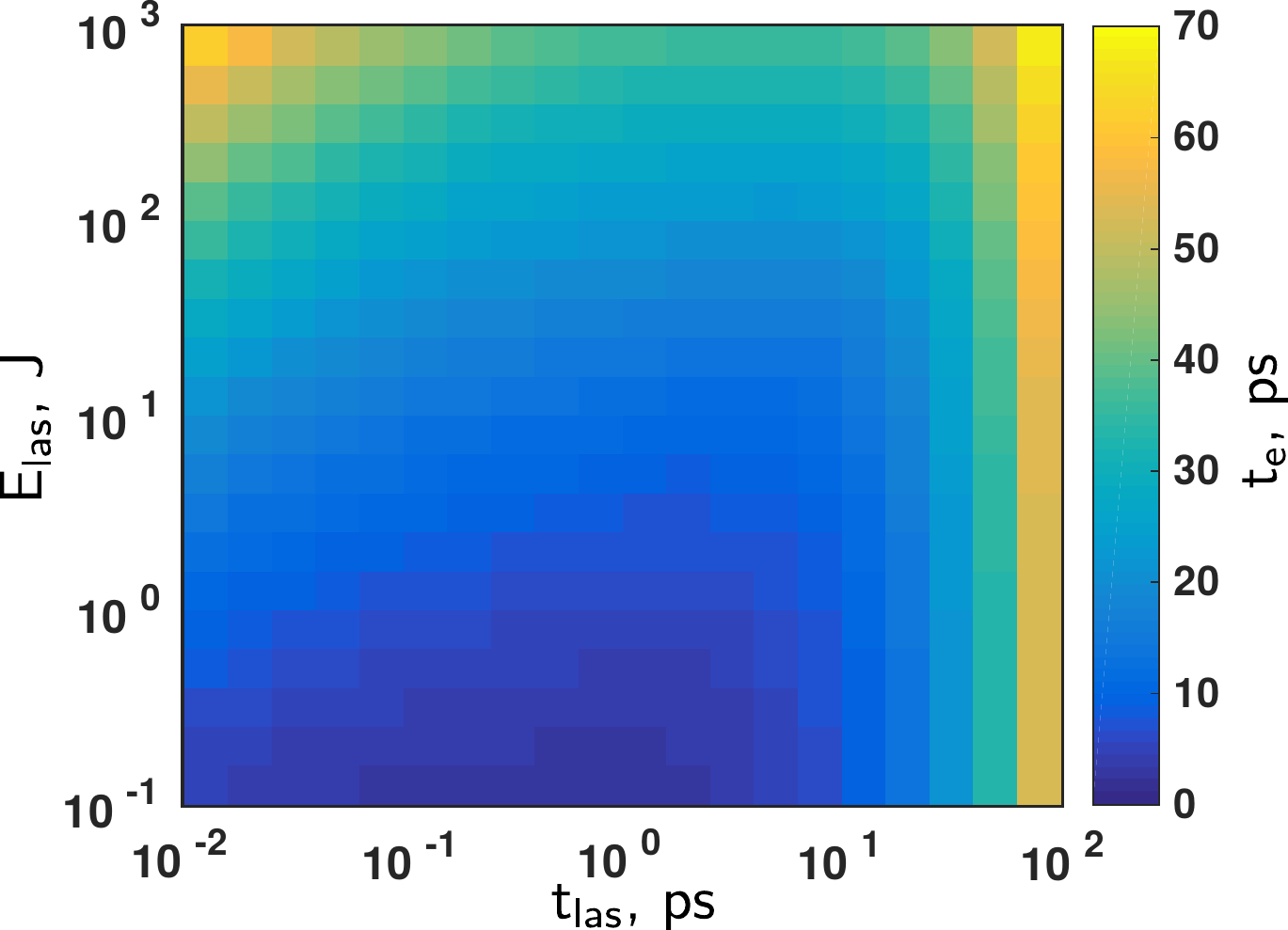}
\includegraphics[width=\columnwidth]{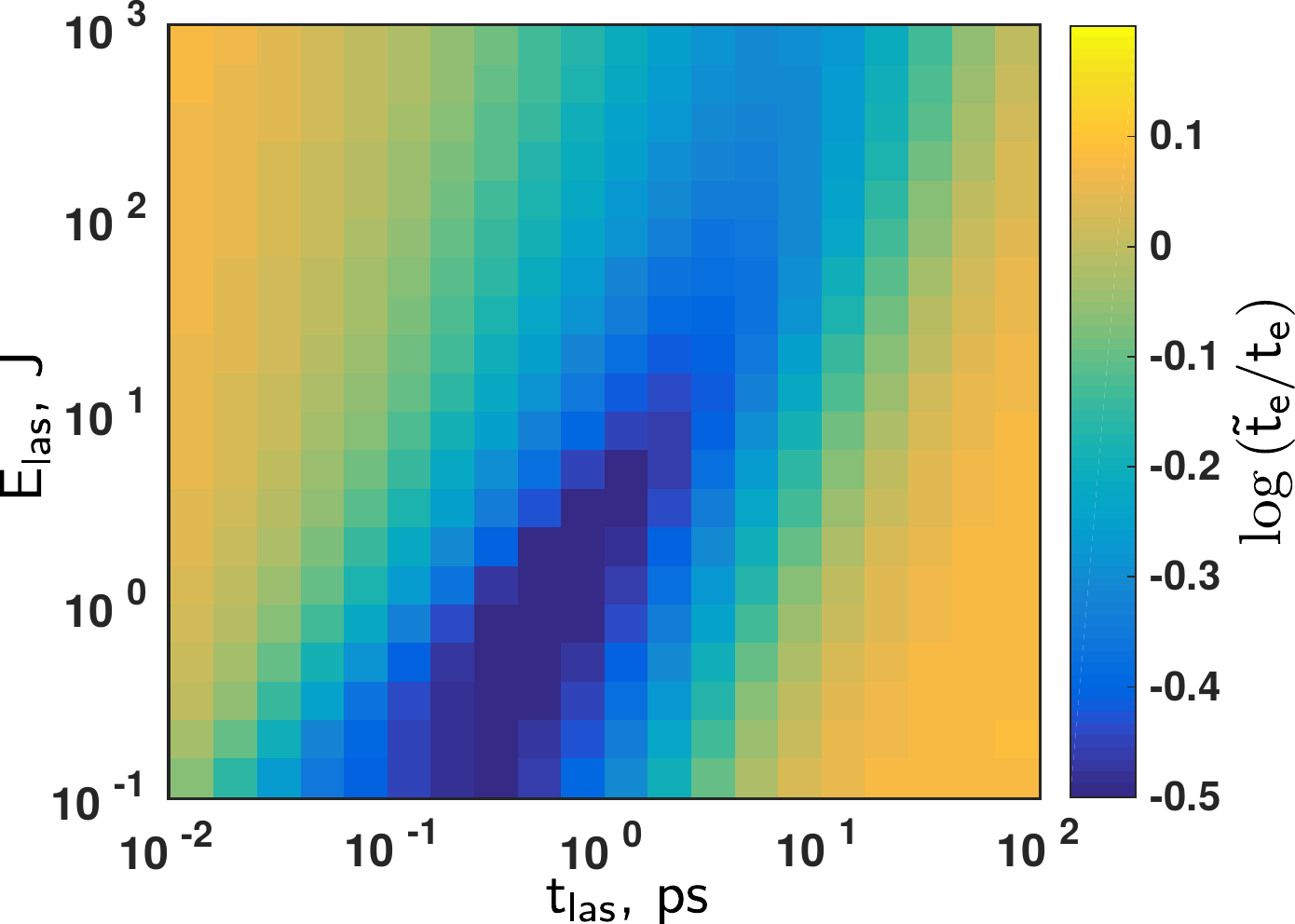}
\caption{\label{fig:ejTime} (Color online) Comparison of two methods of computing the charging time, $t_{\rm e}$, for copper targets. The laser FWHM is $10~\mu$m, with laser absorption $\eta = 40 \%$. The target is assumed infinite. Left plot: Values of $t_{\rm e}$ computed using ChoCoLaTII.f90. The charging time is defined as the moment when the target has accumulated 80\% of its total charge. Right plot: Logarithm of the ratio between Eq.~\eqref{eq:Chargingtime} and ChoCoLaTII.f90 values from the left plot. }
\end{figure*}

\section{Calculation of target charging time}
\label{sec:charging}

To calculate the return current in the target, one needs to estimate the charging time, $t_{\rm e}$: a dipole radiation
model only applies if the charging time is much smaller
than the antenna time ($t_{\rm e} \ll \tau$).
This quantity is difficult to measure because the return current is measured far away from the laser focal spot, where the target holder connects to the ground. In this paper, we make the assumption that the return current near the target is equal to the electron ejection current.

We then consider two regimes in the electron ejection process.\cite{poy15} For high-intensity lasers ($I_{\rm las}>10^{18}$~W/cm$^{2}$, ps-duration or shorter), electrons are ejected during the cooling time of hot electrons generated in the laser-matter interaction. 
For low intensity laser pulses ($I_{\rm las}<10^{17}$~W/cm$^{2}$, ns-duration or longer), electron ejection takes place primarily during the laser pulse. In the case of low laser intensity, the cooling time is much shorter than the laser pulse duration.
To accommodate both regimes, we define the charging time via:
\begin{equation}
    \tilde{t}_{\rm e} = t_{\rm c} + t_{\rm las}  \, , \label{eq:Chargingtime}
\end{equation}
where $t_{\rm c}$ is the hot electron cooling time and $t_{\rm las}$ is the laser pulse duration. 
Generally, to define the cooling time, one needs to calculate the hot electron temperature, $T_0$.\cite{Fabbro1985, Beg1997, Wilks1992, Gibbon2005}
The cooling time is an average over the electron
energy distribution. 
However, the ejection process only lasts as long as there are electrons in the target with energies greater than the target potential.\cite{poy15} Assuming this potential barrier corresponds to an electron temperature $T_0$, we can restrict the cooling time averaging to electrons with energies larger than $T_0$:
\begin{equation}   
 t_{\rm c} (\varepsilon > T_0)= \frac{\exp\left(1\right)}{T_0}\int_{T_0}^\infty  \frac{R(\varepsilon)}{v(\varepsilon)} \, \exp \left( - \frac{\varepsilon}{T_0} \right) \, {\rm d} \varepsilon \, , \label{eq:tc}
\end{equation}
where $R$ is the maximal penetration range of electrons with velocity $v$ in the target. The maximal range can be obtained from the database ESTAR,\cite{estar} or described analytically.\cite{kan72}

To test Eq.~\eqref{eq:Chargingtime}, we can compare it with numerical simulations from the model ChoCoLaTII.f90 which calculates the ejection current. In these simulations, we assume that $t_{\rm e}$ is the time to reach 80\% of the total charge $Q$. This is justified because the ejection current decreases asymptotically: the last 20\% of the ejected electrons take as much time to escape as the previous 80\%.

Figure~\ref{fig:ejTime} compares the analytical and simulated values of $t_{\rm e}$ for copper targets. The left hand graph shows simulations run with ChoCoLaTII.f90, while the right hand graph plots the ratio between $\tilde{t}_{\rm e}$  (calculated using Eq.~\eqref{eq:Chargingtime}--~\eqref{eq:tc}) and $t_{\rm e}$ (from ChoCoLaTII simulations). The left portion of both plots corresponds to the dominance of $t_{\rm c}$ in the ejection time, while the right portion corresponds to the dominance of $t_{\rm las}$. Both estimation methods give the same results in these well-established ejection regimes. Eq.~\eqref{eq:Chargingtime} fails in the middle of the plots, when $t_{\rm c}$ and $t_{\rm las}$ are of comparable size and therefore cannot be neglected.


\section{Conclusion}
\label{sec:conclu}

We have shown that EMP generation is sensitive to
laser and target parameters: in particular the duration of the interaction, the charge accumulated in the target and target geometry. We use a frequency-domain antenna method to estimate the maximum EMP magnetic field, combining calculations of laser-target charging (using ChoCoLaTII.f90) with charge propagation across an antenna-like target. We show that EMP amplitude is independent of laser pulse duration for low energy, short pulse interactions (i.e.\ interactions where the duration of the laser pulse is short compared to the transit time of the return current across a target), whilst longer pulses behave like a constant current source and emit a weaker EMP. 


We extend the description of GHz-frequency EMPs
produced in high power laser experiments to account for
the shape of the target holder. We demonstrate that cylindrical stalks with a large metallic ground, behaving as a dipole antenna like holder A, emit a magnetic field that is proportional to the total charge and to the antenna frequency (i.e. inversely proportional to the length of the stalk). A helix-shaped holder (like C) is found to significantly reduce EMP emissions from the target without altering the laser-matter interaction.
We also show that EMPs are not directly affected by the target charging time (and therefore the laser duration time), provided the charging time is smaller than the characteristic time of the antenna.

We propose a simple procedure for estimating the maximum amplitude of laser-driven EMP. Simulations with the ChoCoLaTII.f90 code provide the charge, $Q$, accumulated on the target. Then the antenna frequency is estimated from the effective holder length or extracted from preparatory shots at low energy. Combining these values using our frequency-domain model, one can estimate the magnetic field amplitude for holders similar to holder A (cylindrical stalk). Conversely, our model allows one to estimate the target charge based on measurements of the magnetic field amplitude. We also provide two methods to estimate the target charging time, which is important for checking the validity of our dipole model.

\acknowledgments

The authors gratefully acknowledge fruitful discussions with Matthieu Bardon, Jakub Cikhardt, Yves Elskens, Josef Kr{\'a}sa and David Neely.
This work was partially done 
when A.~Poy\'e was affiliated to CELIA, University of Bordeaux-CNRS-CEA. The authors would also like to acknowledge funding from EPSRC grants EP/L01663X/1 and EP/L000644/1, the Newton UK grant, the National Natural Science Foundation of China NSFC/11520101003, and the LLNL Academic Partnership in ICF. This research was partially supported by the Project LQ1606 with the financial support of the Ministry of Education, Youth and Sports as part of targeted support from the Czech National Programme of Sustainability II.

\bigskip

The data that support the findings of this study are available from the corresponding author upon reasonable request.

\end{document}